\documentclass[pra,twocolumn,superscriptaddress,floatfix]{revtex4}
\usepackage{hyperref}
\usepackage{graphicx} 
\usepackage{amsmath}
\usepackage[dvips]{epsfig}
\usepackage{color}

\newcommand{\beq}{\begin{equation}}
\newcommand{\eeq}{\end{equation}} 
\newcommand{\beqa}{\begin{eqnarray}}
\newcommand{\eeqa}{\end{eqnarray}}
\newcommand{\ba}{\begin{array}}
\newcommand{\ea}{\end{array}}

\begin{document}
\title{Out-of-equilibrium dynamics of repulsive Fermi gases in quasi-periodic potentials: \\ a Density Functional Theory study}

\author{Francesco Ancilotto}
\affiliation{Dipartimento di Fisica e Astronomia ``Galileo Galilei'', 
                 Universit\`a di Padova, via Marzolo 8, 35122 Padova, Italy}
\affiliation{CNR-IOM Democritos, via Bonomea, 265 - 34136 Trieste, Italy}

\author{Davide Rossini}
\affiliation{Dipartimento di Fisica, Universit\`a di Pisa and INFN, Largo Pontecorvo 3, I-56127 Pisa, Italy}

\author{Sebastiano Pilati}
\affiliation{Dipartimento di Fisica e Astronomia ``Galileo Galilei'', 
                 Universit\`a di Padova, via Marzolo 8, 35122 Padova, Italy}
\affiliation{School of Science and Technology, Physics Division, University of Camerino, 
             Via Madonna delle Carceri 9, I-62032 Camerino, Italy} 

\begin{abstract} 
The dynamics of a one-dimensional two-component Fermi gas in the presence of a quasi-periodic 
optical lattice (OL) is investigated by means of a Density Functional Theory approach.
Inspired by the protocol implemented in recent cold-atom experiments 
---designed to identify the many-body localization transition--- 
we analyze the relaxation of an initially prepared imbalance between the 
occupation number
of odd and of even sites.
For quasi-disorder strength beyond the Anderson localization transition, the imbalance survives 
for long times, indicating the inability of the system to reach local equilibrium. 
The late-time value of the imbalance diminishes for increasing interaction strength.
Close to 
the critical quasi-disorder strength corresponding to the noninteracting (Anderson) transition,
the interacting system displays an extremely slow relaxation dynamics, consistent with sub-diffusive behavior.
The amplitude of the 
imbalance fluctuations
around 
its
running average 
is found to decrease with time, and such damping is more effective
with increasing interaction strengths.
While our study addresses the setup with two equally intense OLs, 
very similar effects due to interactions have been observed also in recent cold-atom 
experiments performed in the tight-binding regime, i.e.
where one of the two OLs is very deep and the other is much weaker.
\end{abstract} 

\date{\today}
\pacs{}
\maketitle

\section{Introduction}
\label{secintro}
Since Anderson's 1958 seminal paper~\cite{anderson}, 
it is known that sufficiently strong disorder can 
cause the localization of noninteracting quantum particles, inducing an insulating behavior 
in  
macroscopic samples.
A vast body of more recent theoretical work supports the view that localization can persist also 
in the presence of interactions~\cite{Basko}, leading to a nonergodic phase 
of matter ---dubbed many-body localized (MBL) phase--- which fails to thermalize, 
thus violating the eigenstate thermalization hypothesis (for a review, see Ref.~\cite{huse}). 
However, the MBL phase is believed to be qualitatively different from the noninteracting Anderson 
insulator: while both phases are characterized by the absence of transport of any physical quantity, 
quantum correlations can still propagate in the MBL phase.
The latter observation comes from the possibility to map the MBL system to an integrable one, with an extensive number of localized constants of motion~\cite{abanin,huse2}: even if local observables reach stationary, and non-thermal, values, the coherences of far apart sites evolve non trivially in time, thus resulting in a slow persistent dephasing ~\cite{abanin3}.
The logarithmic growth in time of the bipartite entanglement entropy~\cite{znidaric, moore, abanin2}, 
or the power-law decay of two-site entanglement~\cite{iemini, pollmann} are two distinctive signatures of that mechanism, 
to be contrasted with a saturating behavior in the Anderson localized phase.

In the last few years cold-atom setups have been employed to experimentally investigate localization phenomena in 
disordered and interacting quantum systems, exploiting the direct control of interactions and 
disorder that these systems allow~\cite{inguscio,bouyer,esslinger,demarco}.
In particular, in a series of experiments reported 
in Refs.~\cite{bloch1,blochgriffiths,blochPRX2D}, the dynamics of a one-dimensional (1D) atomic 
Fermi gas exposed to a quasi-periodic potential has been explored.
The protocol consisted in preparing an initial density modulation, and in studying the 
ensuing
relaxation dynamics for different degrees of (quasi) disorder and interaction strengths. 
The long-time persistence of the initial density imbalance, which signals the failure of a local observable 
to equilibrate, was interpreted as a signature of MBL.

The computational studies of this kind of dynamics usually employ very accurate, but also extremely 
demanding, exact diagonalization or density-matrix renormalization group (DMRG) calculations~\cite{dmrg},
which are limited to small-sized, discrete lattice models and short evolution times.
Such limitations are of particular relevance in the ergodic phase, where the bipartite
entanglement entropy spreads linearly in time, thus making long-time DMRG calculations unfeasible.

In this work we use time-dependent density functional theory (TD-DFT) 
in order to simulate the dynamics of a pseudo-disordered 
two-component 1D Fermi gas with contact repulsive interactions. 
The gas is subject to the quasi-periodic potential generated by two OLs with 
incommensurate periods, but equal intensities (this intensity plays the role of quasi-disorder strength).
Specifically, we analyze the evolution of an initially prepared configuration with a 
density modulation such that the odd sites are empty and the even sites are doubly occupied. 
The exchange-correlation functional, which is the main ingredient of the TD-DFT formalism
and embodies correlation effects beyond the Hartree mean-field description,
is derived using an adiabatic local spin-density approximation (LSDA), 
based on the exact Bethe-Ansatz solution for the homogeneous 
system~\cite{tosi06,abedinpour2007emergence,Xianlong2008} 

The accuracy of this DFT method in predicting ground-state properties has been carefully tested 
in a recent work by making extensive 
comparisons against unbiased quantum Monte Carlo simulations~\cite{pilati}, 
showing that the LSDA approach is very reliable in a broad range of interaction strengths 
and of OL intensities. 
Furthermore, the accuracy of the dynamics 
obtained within the adiabatic approximation for the exchange and correlation functional
has been verified in the context of the 1D Hubbard model by making quantitative 
comparison against essentially exact time-dependent DMRG calculations~\cite{weili}.
We thus argue that the TD-DFT method we employ represents a useful complement with respect 
to more accurate but more demanding methods such
as DMRG calculations, allowing to address larger 
system sizes and longer evolution times.
We remark that the TD-DFT theory in the present formulation
is not restricted to small deviations from the 
ground-state density, but can be reliable well outside the
linear response regime~\cite{ekugross}.

Furthermore, the DFT approach is based on a continuous-space model, 
as opposed to the discrete-lattice tight-biding approximations 
usually addressed by exact diagonalization and DMRG calculations.
This allows us to consider the experimental  setup with two equally intense OLs, which is 
characterized by the presence of mobility edges separating localized from extended 
single particle states~\cite{dassarma1990,boers,biddle,pilatibichro}, in contrast to 
the tight-binding Aubry-Andr\'e model~\cite{aubry} 
---which approximately describes the setup where one OL is very deep and the other is much weaker--- 
for which all eigenstates localize at the same quasi-disorder strength. 
The important effects due to the presence of single-particle mobility edges~\cite{lidassarma} 
beyond the strict tight-binding regime have indeed
been observed and emphasized in a recent experiment~\cite{bloch3}.

In the noninteracting case, our calculations show that the density imbalance 
between odd and even sites rapidly vanishes for weak quasi-disorder, while it survives in the long-time 
limit for quasi-disorder strengths beyond the critical point of the Anderson localization.
In the interacting case, the long-time value of the 
imbalance at strong quasi-disorder is substantially reduced compared 
to the noninteracting case, indicating that interactions have a delocalizing effect. 
Interestingly, when we tune the quasi-disorder strength close to and slightly above the noninteracting Anderson transition, 
the interacting system displays an extremely slow relaxation dynamics, consistent 
with a dynamical critical exponent larger than $z = 2$, thus indicating subdiffusive behavior. 
This phenomenon has been recently observed in experiments~\cite{blochgriffiths} performed in the tight-binding regime, 
and it has been interpreted as a Griffiths effect.
The temporal fluctuations
of the imbalance around its running average are found to decay in time 
and the damping of the fluctuation amplitude 
is found to be more effective with increasing
interaction strengths.
These effects have been observed in previous DMRG simulations of the Aubry-Andr\'e model, 
and have been attributed to the growth of the entanglement entropy~\cite{blochgriffiths}.

The remainder of the article is organized as follows: in Sec.~\ref{secmethods} we provide the details of 
the model we simulate and we describe the TD-DFT computational method used here. 
The results for the relaxation dynamics of the initially imprinted density wave are reported in Sec~\ref{secresults}. 
In Sec.~\ref{secconclusions} we draw our conclusions.

\section{Methods}
\label{secmethods}

The 1D atomic Fermi gas considered in this paper is
described by the following continuous-space Hamiltonian:
\beq
\label{hamiltonian}
\hat {H}=\sum _{i=1}^N \left( -{\hbar ^2\over 2m}{d^2\over dx_i^2}+v(x_i) \right)+
\sum _{i_\uparrow,i_\downarrow}g \,\delta (x_{i\uparrow}-x_{i\downarrow}).
\eeq
Here $N=N_\uparrow + N_\downarrow $, where $N_\uparrow, N_\downarrow$ are
the numbers of atoms in the two
fermionic components, hereafter referred to as 
spin-up and spin-down particles.
The coupling constant $g$ is related to the 1D scattering length $a_{\rm 1D}$, 
$g=-2\hbar^2/(ma_{\rm 1D})$ (with $\hbar$ the reduced 
Planck constant and $m$ the atomic mass).
We consider here purely repulsive interactions, i.e. $g\ge 0$.

The Hamiltonian (\ref{hamiltonian}) faithfully describes cold-atom experiments performed 
in tight cigar-shaped traps, and the value of $a_{\rm 1D}$ can be determined from the  
(three-dimensional) s-wave scattering length and the radial confining strength~\cite{olshanii}.
It is convenient to introduce the adimensional interaction parameter 
$\gamma =mg/(\hbar ^2n)=2/(n|a_{\rm 1D}|)$
where $n=N/L$ is the total density and $L$ the system size.
The $\gamma \rightarrow 0$ ($a_{\rm 1D}\rightarrow -\infty $) limit corresponds 
to a noninteracting Fermi gas, while the $\gamma \rightarrow \infty$ ($a_{\rm 1D}\rightarrow 0^- $) limit 
corresponds to a strongly-interacting regime, where distinguishable fermions 
fermionize~\cite{girardeau1960,girardeau}, i.e., their energy and 
density
can be mapped to those of indistinguishable (spin polarized) fermions~\cite{guan,jochim}.
A 1D (quasi-) disorder can be introduced by superimposing 
two OLs with incommensurate periods, one with (short) period $d_s$, and
another with a longer period $d_l$, thus resulting in an external potential of the form
$v(x)=V_0[\sin^2(\pi x/d_s)+\sin^2(\pi x/d_l)]$.
The OL intensity $V_0$ plays the role of quasi-disorder strength. 
Hereafter $V_0$ will be conveniently expressed in units of 
the recoil energy $E_r=\hbar^2 \pi ^2/(2md_s^2)$ of the short-period lattice.

In order to simulate an infinite quasi-periodic potential, the ratio $d_l/d_s$ between 
the two periods must be an irrational number. We choose the golden ratio
$\phi =(\sqrt{5}+1)/2$ for such number.
Our simulations address a finite box with periodic boundary conditions, since generally they 
reduce finite-size effects compared to, e.g., open boundary conditions.
To make the potential $v(x)$ consistent with the use of periodic boundary conditions,
one needs to approximate this number by the ratio of two integer numbers, the largest one
providing the total length of the periodic cell used in the calculation.
Here we approximate $\phi$ with the ratio of two successive 
numbers in the Fibonacci sequence: $d_l/d_s=F_{k+1}/F_{k}$~\cite{modugno}, 
which converges towards the golden ratio for large values of $k$. 
The potential $v(x)$ thus complies with periodic boundary conditions, 
still being aperiodic within the simulated cell of length $L=F_{k+1}d_s$.
In the following we set (unless otherwise stated) $F_{k}=89$ and $F_{k+1}=144$,
corresponding to a total OL length $L/d_s=144$.
We focus on a half-filled lattice, with $N_\uparrow=N_\downarrow=72$
particles (unless otherwise specified), 
so that on average there is one fermion per well of the short-period lattice. 
It has been recently shown that, in a single half-filled OL,
interparticle interactions play an important role, 
causing the formation of quasi long-range 
antiferromagnetic order~\cite{PhysRevA.96.021601}.
It is also worth emphasizing that a system comprising $N=144$ fermions
cannot be addressed via exact diagonalization calculations 
(see, e.g., the Krylov subspace technique of Ref.~\cite{varma}), 
and is out of reach also for any time-dependent DMRG simulation, 
except 
perhaps in the strongly localized regime, where the entanglement entropy 
does not rapidly grow.

We choose to simulate the dynamics of the Hamiltonian (\ref{hamiltonian}) by employing
a TD-DFT approach. DFT has recently entered the field of ultracold gases
as a useful computational tool that goes beyond the
usual mean-field approximation, which is often used to model such systems.
%
Recent applications of DFT methods to ultracold fermionic systems allowed to study 
the ferromagnetism and antiferromagnetism in repulsive Fermi gases in shallow OLs~\cite{dft}, 
vortex dynamics in superfluid Fermi gases~\cite{bulgac2011real,bulgac2014quantized}, superfluidity 
and density modulations in dipolar Fermi gases~\cite{ancilotto2016kohn},  vortices in rotating 
dipolar Fermi gases~\cite{ancilotto2015kohn}, and the formation of ferromagnetic domains in trapped 
clouds~\cite{zintchenko2016ferromagnetism}.
DFT has also been used to study strongly correlated Fermi gases in elongated harmonic traps~\cite{tosi06}.

In a recent paper~\cite{pilati},  
the accuracy of the LSDA for 1D repulsive Fermi gases in
OLs has been 
assessed. 
To this aim,  
quantum Monte Carlo (QMC) simulations based on 
the fixed-node method to circumvent the sign problem were employed, providing 
exact results for the 1D system of interacting fermions~\cite{ceperley1991fermion}.
A systematic comparison between DFT calculations of 
ground-state energies and density profiles for a
half-filled OL against the outcomes of the QMC 
simulations allowed the authors of Ref.~\cite{pilati} to determine a wide range 
of OL intensities and interaction strengths 
where the LSDA appears to provide quite accurate predictions.
The accuracy of DFT (in the LSDA) in 1D fermionic systems
has been also demonstrated for small finite systems in Ref.~\cite{magyar}.
We note at this point  
that a possible improvement over LSDA for systems
characterized by strongly spatially localized states
(like those arising, e.g., from confinement within deep
optical lattices)
could be the addition of gradient corrections, very much like
to what is currently done in electronic
structure calculations using gradient-corrected 
exchange-correlation Density Functionals~\cite{perdew}. 
%
However, including these corrections does not seem to be necessary here:
the comparisons made in Ref.~\cite{pilati} showed that,
for strengths of the optical lattices similar to
the ones considered here, the LSDA approach already gives results
in excellent agreement with unbiased QMC calculations.

To study the real-time dynamics of the system,
we use here the so-called ``adiabatic'' LSD approximation, 
where the time-dependent exchange-correlation (xc) potential is 
represented by
the static xc potential (treated within the LSD approximation) 
evaluated at the instantaneous density.
The theory is thus local in time, as well as in space
(``memory'' effects are ignored).
An appealing feature of this theory is that it satisfies Galilean invariance~\cite{gross}.
The adiabatic TD-DFT approach (in the LSDA)
to inhomogeneous fermion systems
in 1D has been extensively tested in Ref.~\cite{weili}
and its accuracy in describing collective density and spin dynamics
in strongly correlated 1D ultracold Fermi gases
has been proved by comparing TD-DFT predictions with 
accurate results based on DMRG calculations, 
finding remarkable agreement even beyond the linear response regime.
Including current terms, gradient corrections, or an effective mass might, in principle, 
further increase the accuracy of the TD-DFT approach. These corrections have been employed, e.g, in studies of 
the dynamics of three-dimensional attractive superfluid Fermi gases~\cite{bulgacannualrev}. 
However, in the benchmarks of Ref.~\cite{weili} for repulsive 1D fermions,
excellent agreement has been obtained without including such terms;
therefore, we proceed employing the standard ``adiabatic'' LSDA. 
Notice also that this scheme has been shown to be suitable to account for the dynamics of spin-charge separation, 
a remarkable phenomenon of 1D Fermi systems,
while an additional non-adiabatic term would be needed to account for spin-drag effects~\cite{xianlongPRL2018}.

The Kohn-Sham formulation~\cite{kohn} of DFT~\cite{gross} 
for an inhomogeneous system of $N$ interacting  
particles with spin projection $\sigma =\, \uparrow , \downarrow$ 
is based on the following energy functional of the density:
\beq
E_{\rm KS}[n_\uparrow, n_\downarrow ]={\hbar ^2 \over 2m}\sum _{\sigma } \sum _{i=1}^{N_{\sigma }} 
\int  
|\nabla \phi _i^{\sigma }(x)|^2\, dx +E_{\rm HXC}[n_\uparrow, n_\downarrow ].
\label{k-s}
\eeq
The $\{\phi _i^\sigma (x)\}_{i=1,\ldots, N_\sigma}$ are single-particle orbitals
forming orthonormal sets, $\langle \phi _i^\sigma | \phi _j ^\sigma\rangle=\delta _{ij}$,
filled up to the Fermi level.
The spin-resolved density is given by
$n_\sigma (x) = \sum _{i=1}^{N_{\sigma }}|\phi _i^{\sigma }(x)|^2$,
so that the total density of the system is $n(x)=n_\uparrow(x) + n_\downarrow(x)$.
The interaction energy functional $E_{\rm HXC}$, which includes the mean-field (Hartree)
energy and the exchange-correlation contribution, is treated here within the LSDA, i.e.:
\beq
E_{\rm HXC}=\int dx \: n(x) \: \epsilon _{\rm HXC}^{\rm hom}(n_\uparrow (x),n_\downarrow (x)),
\eeq
where $\epsilon _{\rm HXC}^{\rm hom}$ is the corresponding energy per particle
in the homogeneous phase. The latter can be written 
using the  exact Bethe-Ansatz solution
for the ground-state energy as
\beq
\epsilon ^{\rm hom}_{\rm HXC}=\frac{1}{N} \Big( E_{\rm tot}^{\rm hom}-E_{\rm kin}^{\rm hom} \Big) = {\hbar ^2 \over 2m}n^2f(\gamma, P),
\eeq
where 
\beq
\frac{E_{\rm kin}^{\rm hom}}{N} = \frac{\pi ^2 \hbar ^2 n^2}{24 m}(1+3P^2)
\eeq
is the kinetic energy of the homogeneous non-interacting system, and 
$f(\gamma ,P)=(\pi ^2/4)f_{\rm exa}(\gamma ,P)$. 
Here $P(x)=\big( n_{\uparrow}(x)-n_{\downarrow}(x) \big)/n(x)$ denotes the local polarization.
The term $f_{\rm exa}$ 
is given by \cite{abedinpour2007emergence}:
\begin{eqnarray}
f_{\rm exa}&=&[\eta (x)-1/3]\{1+ \alpha (x)P^2 \nonumber \\
&& +\beta (x)P^4  - [1+\alpha (x)+ \beta (x)]P^6 \} ,
\end{eqnarray}
where $x\equiv 2\gamma/\pi $ and:
\begin{eqnarray}
  \alpha (x) & = & {(-x^2 +a_\alpha x +b_\alpha)\over x^2 + c_\alpha x -b_\alpha} , \\
  \beta (x)  & = & {a_\beta(x) \over x^2 +b_\beta x +c_\beta} , \\
  \eta (x)   & = & {4x^2/3 +a_P x +b_P \over x ^2+c_Px +d_P } .
\end{eqnarray}
Here: 
$a_{\alpha }=-1.68894$,
$b_{\alpha }=-8.0155$,
$c_{\alpha }=2.74347$,
$a_{\beta }=-1.51457$,
$b_{\beta }=2.59864$, 
$c_{\beta }=6.58046$,
$a_P=5.780126$,
$b_P=-(8/9)\ln 2+\pi a_P/4$,
$c_P=(8/\pi)\ln 2+3a_P/4$,
$d_p=3b_P$.

Constrained minimization of the functional $E_{\rm KS}$
leads to the coupled KS eigenvalues equations:
\beq
 \label{kseq}
  \hat {H}_{\rm KS}\,\phi _i^\sigma (x)\equiv \! \left[-{\hbar ^2 \over 2m }{d^2\over dx^2} \!+\! v(x) \!+\!V_{\sigma}(x) \right] \!\phi _i^\sigma (x)=
\epsilon _i \phi _i^\sigma (x).
\eeq
The effective potential 
$V_{\sigma}(x) \equiv \delta E_{HXC} / \delta n_{\sigma}(x) 
=
\partial (n\epsilon _{HXC}) / \partial n_{\sigma}$ 
can be written as
\begin{eqnarray}
V_{\sigma} & = & { \hbar ^2\over 2m}\left[ f(\gamma ,P) \frac{\partial n^3}{\partial n_{\sigma}} + n^3 \frac{\partial f}{\partial n_{\sigma}} \right]
\nonumber \\
& = & 
{ \hbar ^2\over 2m}  \left[  3n^2 f-n^2\gamma {\partial f\over \partial \gamma} \pm
2n n_{-\sigma} \, \frac{\partial f}{\partial |P|} \, {P\over |P|}  \right],
\label{veff}
\end{eqnarray}
where we used the fact that $\partial n/\partial n_{\sigma}=1$, $\partial P/\partial n_{\sigma}=\pm 2n_{-\sigma}/n^2$,
and $\partial \gamma /\partial n=-\gamma/n$.
Therefore $V_{\sigma}$ couples only fermions with opposite polarization, 
since we consider a zero range model for the interatomic interaction.

In the following we seek for time-dependent 
solutions $\{ \phi _i ^\sigma (x,t)\}_{i=1,\ldots,N_\sigma}$
by propagating in real time the time-dependent version~\cite{kohn} of the KS equations (\ref{kseq}), i.e. 
$i\hbar \partial \phi _i^\sigma /\partial t=\hat{H}_{\rm KS}\phi_i^\sigma $.
Both the densities $n_\sigma(x)$ and the orbitals $\phi _i^\sigma(x)$ are
discretized in Cartesian coordinates using a spatial grid fine enough to guarantee
well converged values of the total energy $E_{\rm KS}$. 
The orthogonality between different orbitals is enforced by a Gram-Schmidt process. 
The spatial derivative entering Eq.~\eqref{kseq} is calculated with accurate 13-point formulas.
The time-dependent Schr\"odinger's equation~\eqref{kseq} is solved using 
an Hamming's predictor-modifier-corrector method~\cite{Ral60}, initiated by 
a fourth-order Runge-Kutta-Gill algorithm~\cite{Ral60,Pre92}. 
This choice provides excellent stability and energy conservation even during simulations 
spanning rather long time intervals.

\section{Results}
\label{secresults}
\begin{figure}[t]
\centerline{\includegraphics[width=1.0\linewidth,clip]{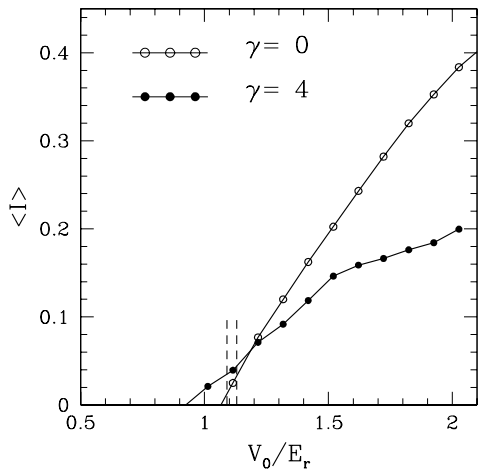}}
\caption{
Late-time value  of the imbalance $\langle{\cal I}\rangle$ as a function of 
the quasi-disorder strength $V_0/E_r$, computed as the average of ${\cal I}(t)$ 
within the time window $t \in \left[80\tau:100\tau\right]$ after the relaxation starts.
We choose $\tau=2md_s^2/\hbar$ as the time unit.
Two cases are shown:
the non-interacting ($\gamma = 0$) and the interacting ($\gamma =4$) case.
In the former case, the dynamical evolution 
of ${\cal I}(t)$ rapidly saturates, so that the values of $\langle{\cal I}\rangle$ shown here represent the 
asymptotic stationary value. In the latter case, ${\cal I}(t)$ undergoes an extremely slow drift if $V_0/E_r \gtrsim 1.1$
(see Fig.~\ref{fig3}), 
making it unfeasible to identify an asymptotic stationary value within the achievable simulation times.
The vertical dashed segments bracket the critical point of the noninteracting 
Anderson localization transition, computed by analyzing the participation ratios 
of the low-energy single-particle orbitals (see text).\\
}
\label{fig1}
\end{figure}

In order to discern the delocalized ergodic phase from the insulating (putative MBL) phase,
we follow a protocol similar to the one used 
in  a series of recent experiments~\cite{bloch1,blochgriffiths,blochPRX2D,bloch3}.
We create an initial state with a density modulation, such that the even sites of the short-period OL
are almost empty and the odd sites are almost doubly occupied. 
This is achieved by computing the ground state of the Hamiltonian (\ref{hamiltonian}) in the presence 
of an additional superimposed OL with period $2d_s$ and a
well depth which is twice the chosen value of $V_0$ \cite{footnote}.
The dynamics of this initial state is determined via the TD-DFT method described in Section~\ref{secmethods}. 
In particular, we compute the time dependent imbalance
${\cal I}(t)$ between the respective atom number on even, $N_e$, and odd, $N_o$, sites:
\beq
{\cal I}(t)= {N_e-N_o\over N_e+N_o}.
\label{imbal}
\eeq
In the noninteracting case, the imbalance ${\cal I}$ rapidly 
reaches negligibly small values for quasi-disorder strengths
smaller than $V_0 \simeq 1.06 E_r$, as shown in Fig.~(\ref{fig1}),
indicating that the system is indeed able to equilibrate.
For higher values of the quasi-disorder strength $V_0$, 
${\cal I}$ remains finite in the long time limit.  
Its asymptotic value $\langle{\cal I}\rangle$ (computed as described below)
increases with $V_0$ for disorder strengths above the critical point. 

The position of the calculated critical 
point for the noninteracting system
is consistent with the quasi-disorder strength necessary 
to induce Anderson localization of the 
single-particle eigenstates in the low-energy regime of the spectrum,
equal to $V_0 \simeq 1.1 E_r$.
We determine this value by analyzing 
the scaling with system size of the average of the participation ratios 
(which is a measure of the spatial extent of a single-particle wavefunction~\cite{kramer}) 
of the lowest $L/(60d_s)$ eigenstates  
(the vertical segments shown in Fig.~\ref{fig1} bracket the so-determined critical point).
This suggests that, as soon as some of the single-particle eigenstates are spatially localized, 
the asymptotic value of $\langle {\cal I} \rangle$ is finite.
It is worth emphasizing that, as opposed to the 
Aubry-Andr\'e model ---for which all eigenstates localize 
at the same quasi-disorder strength--- in the continuous-space model 
of Eq.~(\ref{hamiltonian}) the critical quasi-disorder strength
depends on the energy of the state~\cite{dassarma1990,boers,biddle,PhysRevA.96.021601}. 
In particular, the low energy eigenstates localize at weaker quasi-disorder strength compared to high energy states.

\begin{figure}[t]
\centerline{\includegraphics[width=1.0\linewidth,clip]{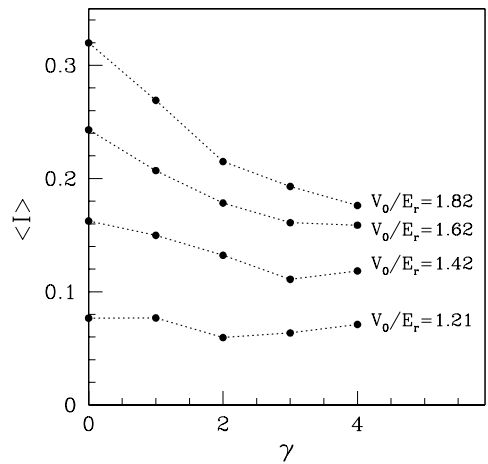}}
\caption{
Late-time average value of the imbalance $\langle{\cal I}\rangle$, for various values of the 
quasi-disorder strength $V_0/E_r$, shown as a function of the interaction strength $\gamma$.
}
\label{fig2}
\end{figure}

%

\begin{figure}[t]
\centerline{\includegraphics[width=1.0\linewidth,clip]{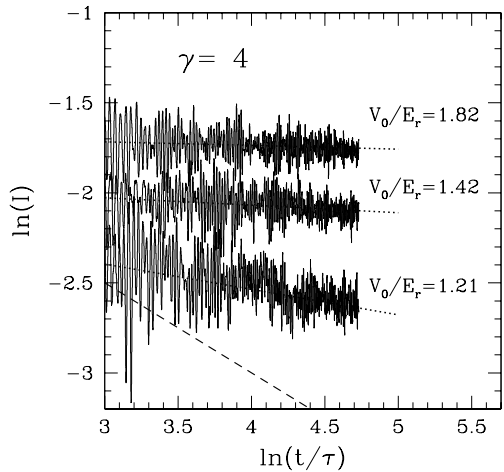}}
\caption{
Imbalance ${\mathcal I}$ as a function of evolution time $t$ for the 
interacting gas at $\gamma = 4$, in three bichromatic OLs with 
intensities (from bottom to top) $V_0/E_r=1.21, 1.42, 1.82$.
The dotted lines represent power-law fits of the type ${\mathcal I} \sim t^{-1/z}$, where $z$ is 
the dynamical critical exponent. The dashed line corresponds to $z=2$, 
which characterizes a purely diffusive behavior.}
\label{fig3}
\end{figure}

Introducing interactions among fermions (i.e. $\gamma >0$) causes 
important effects. For strong quasi-disorder, 
the late-time value of ${\cal I}$ is 
significantly reduced compared to the noninteracting case, 
while for intermediate quasi-disorder strengths this reduction is less pronounced.
This is apparent in Fig.~\ref{fig2}, where we plot 
the late-time average value $\langle{\cal I}\rangle$ versus interaction strength $\gamma$ 
for different quasi-disorder intensities $V_0$.  
The values of $\langle{\cal I}\rangle$  
displayed in  Fig.~\ref{fig1} and Fig.~\ref{fig2} are computed by averaging the 
calculated values ${\cal I}(t)$
over the last portion, $\sim 20\,\tau$, of a total simulation time 
$t_{\rm max}\sim 100\,\tau $. 
Here $\tau\equiv 2md_s^2/\hbar$ is the unit of time.
%
%
%
A word of caution is in order here.
The time evolutions of the imbalance ${\cal I}$ in the noninteracting case and 
in the interacting case are qualitatively different.
While in the former case 
${\cal I}(t)$ rapidly saturates to the asymptotic value, 
and then undergoes virtually random fluctuations around the mean value,
in the latter case, if the quasi-disorder strength is close to or slightly beyond 
the critical point of the (noninteracting) 
Anderson transition, we observe an extremely slow drift 
towards lower values, i.e. longer relaxation times. 
%
Therefore, in the interacting case the late-time average imbalance $\langle {\cal I} \rangle$ measured
at quasi-disorder strengths close to the Anderson transition should not be interpreted
as an asymptotic stationary value, but rather as a transient value observed at an intermediate
time along an extremely slow relaxation dynamics.
%
%
This slow relaxation is illustrated in Fig.~\ref{fig3}, where we show the different dynamics 
associated with various interaction strengths.
A similar slowdown of the dynamics has previously 
been observed in recent experiments~\cite{blochgriffiths}, 
and also in exact-diagonalization 
calculations~\cite{blochgriffiths} (see also Refs.~\cite{mondaini,sierant}), 
both performed in the tight-binding regime,
and it was interpreted as a consequence of 
the Griffiths effect. This effect is characteristic of 
purely random systems, where statistical spatial variations of the 
external random field create subregions with stronger 
disorder, which have a local insulating character. 
While such subregions will eventually thermalize with the 
surrounding (thermal) regions, they cause a slowdown of the overall dynamics. 
The existence of the Griffiths phenomenon for quasi-periodic 
systems has been challenged~\cite{knap}, due to 
the absence of purely random statistical fluctuations. 
In Ref.~\cite{blochgriffiths}, the occurrence of Griffiths 
effects in the quasi-periodic system was attributed 
to the randomness of the initial state, in which the spin distribution was disordered, 
causing a different local impact 
of the interactions. It is indeed remarkable that a similar effect 
is observed also in our study, where the initial state is 
instead 
ordered
(i.e., an alternation of almost empty and doubly occupied sites).

\begin{figure}[t]
\centerline{\includegraphics[width=1.0\linewidth,clip]{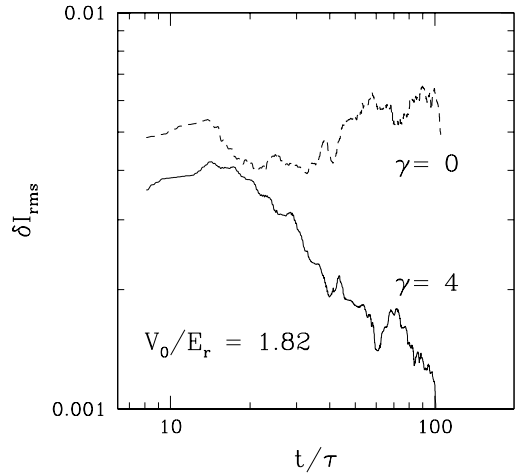}}
\caption{
Average fluctuation of the imbalance around its running average,
as a function of time, for two values of the 
interaction strength $\gamma$. The quasi-disorder strength is $V_0/E_r=1.82$.
}
\label{fig4}
\end{figure}

Following the theoretical analysis of Ref.~\cite{blochgriffiths}, we
fit the decay with time of the imbalance with 
the power law ${\cal I}\sim t^{-1/z}$, where
the dynamical critical exponent $z$ associated to 
transport is used as a fitting parameter. 
In the quasi-disorder range $1.1\lesssim V_0/E_r\lesssim 1.4$ we 
obtain values larger than $z = 2$ (which would 
correspond to diffusive dynamics), thus
indicating subdiffusive behavior. 
For larger quasi-disorder strength, the imbalance (after an initial rapid decay) remains essentially 
constant for the observable timescale, consistently with the emergence of a (putative MBL) 
phase which fails to equilibrate.
It is worth emphasizing that the total time-scale of our simulations is comparable to the longest evolution times achieved in the recent cold-atom experiments, and it is two orders of magnitude longer than the microscopic time-scale for single-particle tunneling between nearest-neighbor wells of a single OL with intensity $V_0\sim E_r$. The recent cold-atom experiments employed deeper OLs, where the tunneling time is somewhat longer than the one corresponding to our setup, so that the total observable time-scale was approximately $40$ times longer than the tunneling time. It is remarkable that memory of the initial configuration survives for times so much longer than the single-particle tunneling time.

 \begin{figure}[t]
\centerline{\includegraphics[width=1.0\linewidth,clip]{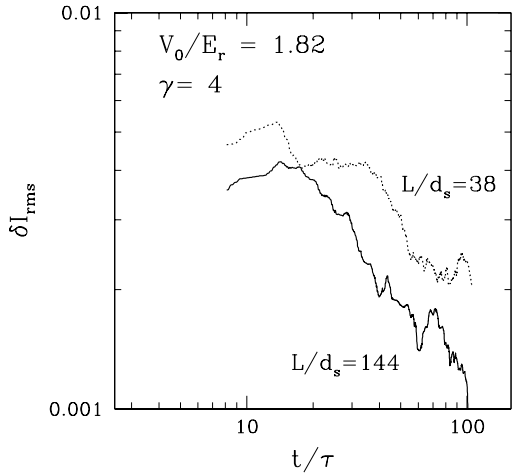}}
\caption{
Average fluctuation of the imbalance around its running average, 
as a function of time, for the case $V_0/E_r=1.82$ and two system sizes $L/d_s$.
Dashed line: $L/d_s=38$; solid line: $L/d_s=144$.
}
\label{fig5}
\end{figure}

Interactions have a relevant impact also on the temporal fluctuations of the imbalance. 
In order to elucidate this effect, we characterize the 
amplitude of these fluctuations using the root mean squared 
deviation $\delta {\cal I}_{\textrm{rms}}$ around the running average
$\langle{\cal I}\rangle_{\textrm{run}}$, evaluated 
within a temporal window of width $\Delta t\sim t_{\rm max}/7$. 
Here 
\beq
\delta {\cal I}_{\textrm{rms}} \equiv \sqrt{ \frac{1}{N_s} 
\sum_{j=1}^{N_s} \Big( {\cal I}(t_j)-\left<{\cal I}\right>_{\textrm{run}} \Big)^2 }
\label{fluct}
\eeq
where the sum and the running average $\langle \dots\rangle_{\textrm{run}}$ are 
performed over $N_s$ time-steps $t_j$ $(j=1,\ldots,N_s)$ within the temporal window.
When the interaction strength increases, the asymptotic value of $\delta {\cal I}_{\textrm{rms}}$ 
diminishes substantially compared to the noninteracting case, 
where the imbalance fluctuations appear 
instead 
to undergo virtually random oscillations, after an initial decay. This is illustrated in Fig.~\ref{fig4}. 
This interaction effect has been discussed previously in Ref.~\cite{bloch1} based on time 
dependent DMRG simulations of the Aubry-Andr\'e model. It was also found that the rate of suppression 
of $\delta {\cal I}_{\textrm{rms}}$ with time is related to the growth rate of the entanglement entropy, suggesting 
that  measuring the fluctuations of ${\cal I}$ might allow one to extract information about the 
entanglement entropy ---a nonlocal quantity--- from a local observable.\\
It is worth noticing that the finite oscillation amplitude $\delta {\cal I}_{\textrm{rms}}$ we measure in the long time 
limit in the interacting system might be due to the finite system size. Indeed, as shown in Fig.~\ref{fig5}, 
the asymptotic value is significantly smaller for larger system sizes.

\section{Conclusions}
\label{secconclusions}

We have studied the dynamics of a two-component 1D
Fermi gas with contact repulsive interactions, and subject to a quasi-periodic potential 
formed by two OLs with incommensurate periods. The setup we considered, in which
the two OLs have the same intensity, has been addressed before only via equilibrium ground-state 
QMC simulations, which allow to discern the metal-insulator transition 
at zero temperature~\cite{pilatibichro}. 
In this article we extended the previous study by 
addressing the out-of-equilibrium dynamics via the TD-DFT method,
following a protocol similar to the one implemented in a
series of recent 
cold-atom experiments~\cite{bloch1,blochgriffiths,blochPRX2D,bloch3} 
aimed at investigating the MBL phenomenon. 
This protocol consists in following the relaxation of 
an initially imprinted density imbalance, and allowed the experimentalists to identify a 
nonergodic phase where the initially 
imprinted density imbalance survives after long times, thus signaling the 
inability of the system to reach local thermal equilibrium. 
This is one of the features characterizing MBL phases~\cite{huse}.

Our simulations displayed several of the most 
relevant phenomena observed in the experiments, which 
however have been performed in the tight-binding setup, 
where one of the two OL is very deep and the other is 
much weaker. 
Among other effects, 
we observed a sizable reduction of the long-time value of the imbalance 
---which is finite in the strong quasi-disorder regime---
due to weak and intermediate repulsive 
interactions. These results represent a quantitative benchmark 
which might be useful for
future experiments performed beyond the tight-binding regime.
Furthermore, we observed an extreme slowdown of the dynamics of the interacting system in the vicinity 
of the 
noninteracting (Anderson) transition, and also a decrease in time of the imbalance fluctuations. 
This decrease  is quite pronounced in the interacting system, while it is essentially negligible in the noninteracting case.
We 
underline
that the continuous-space model we 
consider here differs substantially from the Aubry-Andr\'e model 
(which approximates the experimental system in the tight-binding regime).
For this reason we do not observe the reentrant 
behavior observed in the experiment, where the late-time value of the imbalance was found to increase 
in the strongly interacting limit.
This reentrance is due the fact that, in this limit, the dynamics can be described using a 
noninteracting fermion model~\cite{bloch1}. While in the Aubry-Andr\'e model all single-particle states 
---which determine the dynamics of the noninteracting model--- 
localize at the same quasi-disorder strength, in the continuous-space model~(\ref{hamiltonian}) 
high-energy extended states are present also at strong quasi-disorder. Furthermore, in the latter model atoms in doubly occupied sites 
(referred to as doublons in Ref.~\cite{bloch1}) can separate also in the strongly interacting limit, while in the Aubry-Andr\'e model
they become, in this limit, stable quasiparticles which tunnel only with an effective second-order tunneling, henceforth favoring localization.

The TD-DFT method implemented here represents a useful complement 
to 
more accurate but more demanding techniques such as, e.g., DMRG calculations.
Indeed, it allows to address larger systems sizes and 
longer evolution times, and also to simulate realistic continuous-space models as 
opposed to tight-binding approximations. 
Furthermore, TD-DFT can be extended to higher dimensions 
at an affordable computational price.
In contrast, DMRG has revealed a powerful method to address the ground state 
of ladder and even 2D systems (being unbiased with respect to any sign problem)~\cite{dmrg2D}, 
but its cost remains exponential in the system width, and 
its extension to time-dependent calculations 
is problematic, and it is still the subject 
of on-going research in the tensor-network community.

\acknowledgments

We acknowledge the CINECA award under the ISCRA initiative, for the availability of high performance computing resources and support.
S. P. and F. A. acknowledge financial support from the BIRD 2016 project ``Superfluid properties of Fermi gases in optical potentials'' of the University of Padova.
Fruitful discussions with R. Fazio are acknowledged.


\begin{thebibliography}{99}
\bibitem {footnote}
  With such choice the initial occupation of odd sites is between $1.8$ and $2$,
  while that of the even sites is between $0$ and $0.15$.
  A more ordered initial state can be prepared by choosing a larger 
  value for the well
  depth of the additional confining potential, e.g. three times the chosen value of $V_0$.
  However, we verified that the results of our time-dependent simulations 
  are independent of the fine details of the initial state.
\end{thebibliography}
\end{document}